\documentstyle[a4p,12pt,cite,epsfig]{article}
\newcommand{\lll}  {\mbox{\LARGE /}}

\def\beq{\begin{equation}}  
\def\eeq{\end{equation}}
\def\bea{\begin{eqnarray}}
\def\eea{\end{eqnarray}}

\begin{document}
\begin{titlepage}
\vspace{4mm}
\bigskip
\begin{flushright}  
TAUP-2758-04
\end{flushright}
\begin{center}
\vspace{7mm}

\bigskip 



{\Large\bf The particle emitter size dependence on energy\\ 
\vspace{3mm}

in ${\bf e^+e^- }$ annihilations into hadrons} 

\vspace{22mm}

{\Large Gideon Alexander}\footnote{{\it E-mail address:} alex@lep1.tau.ac.il}
\end{center}  
\vspace{2mm}

\centering{\large\it School of Physics and Astronomy}\\
\centering{\it Raymond and Beverly Sackler Faculty of Exact Sciences}\\
\centering{\large\it Tel-Aviv University, 69978 Tel-Aviv, Israel}\\

\vspace{15mm} 

\begin{abstract}
The nearly energy independent hadron emitter dimension $r$, measured 
in $e^+e^-$ annihilation in the energy range 10 to 91 GeV 
via the Bose-Einstein correlation of two identical charged pions,
is shown to be well accounted for by choosing   
the hadron jets as independent pion sources. 
To this end the known normalised  factorial
cumulant moments dependence on particle sources is
adapted to
the Bose-Einstein correlation formalism  to yield a relation between $r$ and
these sources. 
This approach is also able to account for the measured $r$ values
obtained for the $Z^0$ decays into two and three 
hadron jets. Finally the estimated $r$ value of the 
hadronic $\Upsilon(9.46)$ decay via three
gluons is expected to be higher by about 6 to 11$\%$ over that 
predicted for its one photon hadronic decay mode. 

\end{abstract}
\vspace{3mm}
(Submitted to Phys. Lett. B)\\
\centering\today\\
\vspace{2.5cm}
\begin{flushleft}
\end{flushleft}
\vspace{3mm}

\end{titlepage}
\section{Introduction}
\label{intro}
\ \ \ \ \
The two-particle intensity interferometry has been 
proposed by Hanbury-Brown and Twiss \cite{hbt} to
correlate radio waves arriving from outer space for the purpose
of measuring the angular diameter of astronomical objects.
This method, which here is referred to as Bose-Einstein correlations
(BEC),
has been further extended by Goldhaber et al., \cite{gold} 
to the system of identical hadron pairs produced in two-particle collisions. 
Taking the hadron emitter to be a sphere 
of radius $r$ with a Gaussian distribution, the
correlation function $C(p_1,p_2)$ 
of pairs of identical bosons with momenta 
$p_1$ and $p_2$ can be parametrised by
\begin{equation}
C(Q)\ =\ 1\ +\ \lambda e^{-r^2Q^2}\ ,
\label{bec}
\end{equation}
where $Q^2\ =\ -(p_1 - p_2)^2$.  
The $\lambda$ parameter, usually referred to as
the chaoticity parameter, is a measure of the  
incoherence level of the hadrons source. 
However, the estimated $\lambda$ values obtained from 
BEC analyses
are also affected by the purity level of the identical hadron pairs
sample taken from the experimental data.  
In recent years the BEC interferometry of
pairs of charged pions
is extensively used in order to investigate the underlying dynamical
processes occurring in heavy ion collisions at ultra high energies
\cite{heavy}. 
In parallel,
the BEC analysis was and is also applied to $e^+e^-$ annihilations 
at different 
centre-of-mass energies $\sqrt{s}$, and as a function of the outgoing 
hadron mass,
in order to estimate the emitter dimension \cite{rpp}.
Lately a special attention has been given to the role, if any, 
of these BEC
in the reaction $e^+e^- \to W^+W^- \to$ hadrons in order to estimate
their possible effect on the $W$ mass determination \cite{wwcomb}.\\
 
\begin{figure}[ht]
\centering{\psfig{file=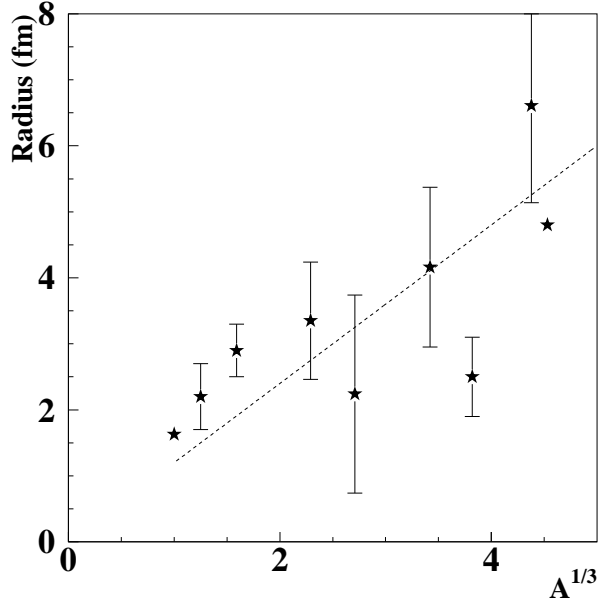,height=8cm, 
bbllx=32pt,bblly=156pt,bburx=532pt,bbury=664pt,clip=}}
\caption{\small A compilation of $r_{\pi^{\pm}\pi^{\pm}}$ extracted from BEC
analyses of identical charged pion-pairs produced in heavy ion
collisions
shown as a function of $A^{1/3}$ where $A$ is the atomic number of
the projectile. The values are taken from Ref. \cite{chacon} and
whenever more than one value was given for the same projectile the
plotted value is their average. The line represents
the relation $r=1.2A^{1/3}$ fm. 
}
\label{comp_nuclei} 
\end{figure}

\begin{figure}[ht]
\centering{\psfig{file=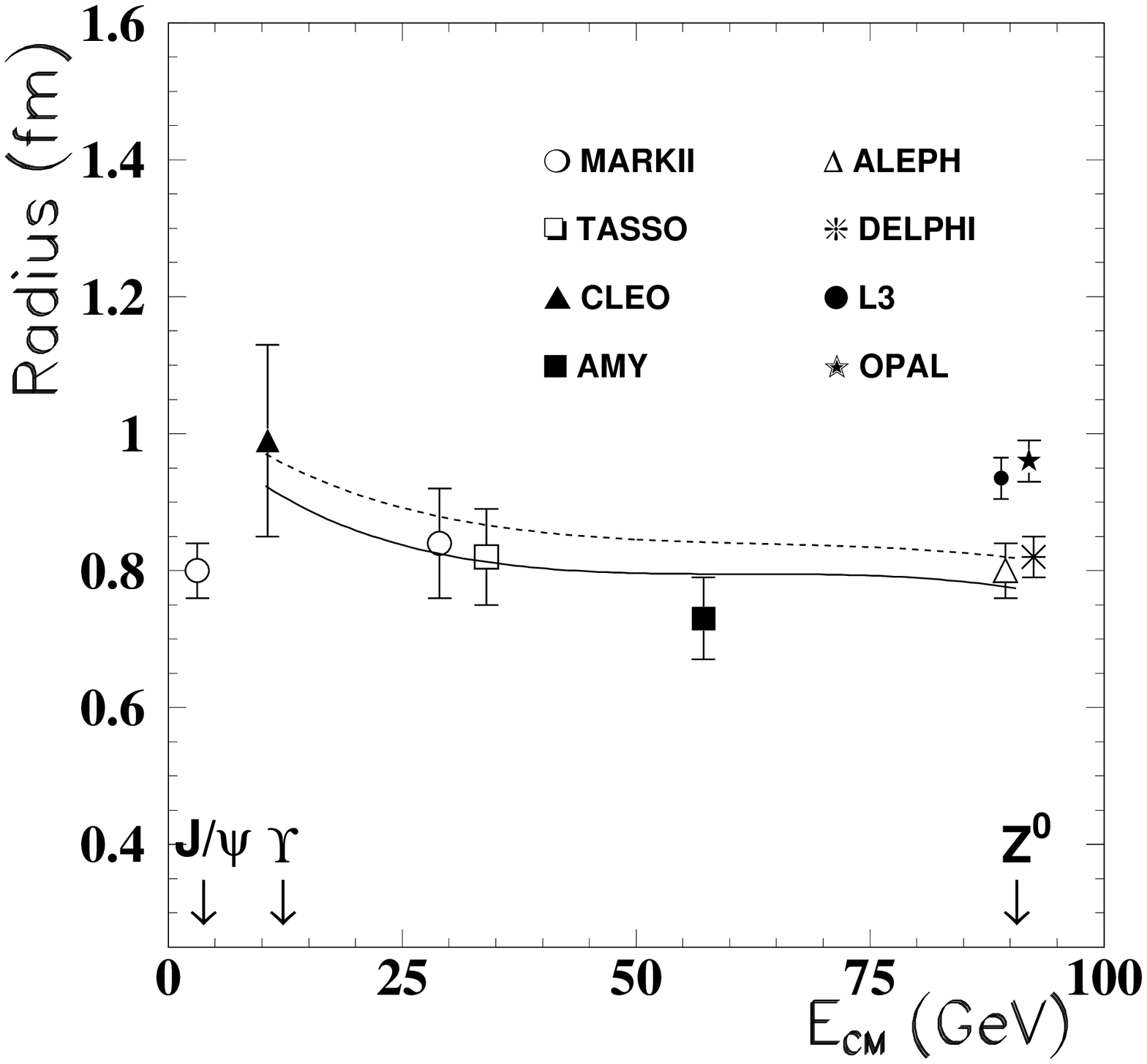,height=8.5cm}}\ \ \ 
\caption{\small
Values of $r_{\pi^{\pm}\pi^{\pm}}$ obtained from Bose-Einstein
correlation analyses of hadrons produced in
$e^+e^-$ annihilations \cite{rpp,l3r}) plotted against
$\sqrt{s}$. 
The shown band is the 
$r_{\pi^{\pm}\pi^{\pm}}$ behaviour expected for 
a constant $\lambda$ value
from the hadron source approach,
normalised at 
91 GeV to the $r$ value obtained by the ALEPH 
collaboration. 
The continuous and dashed lines correspond respectively 
to the $x_{max}$ values of 0.99 and 0.95 used in calculating  
$\langle D_{qqg}\rangle$ via
Eq. (\ref{average}).
}
\label{compilation} 
\end{figure}

An early compilation of the $r_{\pi^{\pm}\pi^{\pm}}$
values measured in heavy ions collisions 
\cite{chacon} is seen in Fig.~\ref{comp_nuclei} to be
rather well described by 
the straight line representing the relation $r\ =\ 1.2A^{1/3}$ fm, 
the known nucleus radius dependence on the atomic number $A$. 
This increase is attributed to the rise of the number of hadron
sources, such as the nucleon-nucleon interaction, which occur as 
the atomic number of the colliding ions increases \cite{heavy}. At the same
time the BEC deduced $r$ values, extracted from $e^+e^-$ annihilations 
leading mainly to pions, seem
to be rather independent of the centre of mass energy
of the colliding electrons. This is illustrated in Fig.~\ref{compilation} 
where the
small deviations from a constant $r$ value are attributed to the
variation in the experimental procedures adopted in each of
the BEC analyses \cite{rpp,l3r}. Additional information concerning the
properties of the BEC extracted $r$ dimension is coming 
from the hadronic $Z^0$ decays where it is found that the $r$ value
increases with number of hadron jets and charged pion
multiplicity \cite{opalmul}.\\
    
Here it is shown  that the observed energy independent  
behaviour of $r$ in $e^+e^- \to$ hadrons annihilations 
is consistent with the assumption that only hadrons emerging from
the same final state hadron jet
can be correlated.
In addition, this approach leads also to an $r$ 
increase of some $10\%$ of the three-jet hadronic
$Z^0$ decays over that expected for two-jets decay events.  
Finally the hadron jet
source approach is applied to the hadronic $\Upsilon(9.46)$ decays
via the three gluons and one photon processes.
\section{The hadron source approach in cumulants and BEC}
\vspace{3mm}

A direct way to extract the genuine dynamical correlations of 
two and more hadrons, 
produced in high energy reactions, is offered by the
so called bin-averaged normalised factorial cumulant moments $K_q$,
here denoted simply as
cumulants, where $q$ is the size of the hadron group and is equal 
e.g. to 2 for a pion-pair system. 
This method, which was first proposed in \cite{first} and
described in details in \cite{dremin}, 
has been applied in the study of multi-particle dynamical
fluctuations. 
In these investigations, also referred to as intermittency analyses, 
the dependence of the cumulants values on the
average multiplicity has been proposed to be the
consequence of a mixing of several well defined hadron emission sources
\cite{emu,lipa,barshay,busch,sark}.\\ 

In this emission source approach, correlations exist only
between particles emitted from the same source so that
the correlation strength is reduced whenever particles are grouped from more
than one source. For a quantitative evaluation of this reduction
we follow here closely 
Ref. \cite{sark} in its treatment of the one dimensional
(rapidity) cumulants $K^{av}_q$, which are averaged over $M$ bins 
with the condition that $M \geq 10$. If one denotes the average value of
a cumulant determined from one hadron source as $K^{av(1)}_q$ then
the cumulant $K^{av(k)}_q$, calculated for all possible q-groups 
from a state $k$
of more than one source,  will be diluted by a factor 
$D^{(k)}_q$ so that
\begin{equation}
K^{av(k)}_q\ =\ D^{(k)}_q ~K^{av(1)}_q\ ,
\label{definition}
\end{equation}
where $D^{(k)}_q$ is equal to
\begin{equation}
D^{(k)}_q ~= ~\frac{P^G_q}{(P^G_q ~+ ~P^{NG}_q)} ~.
\label{dilution}
\end{equation}
Here $P^G_q$ denotes the number of q-particle groups, e.g. pairs of pions,
emerging from the same source while
$P^{NG}_q$ stands for the number of all possible combinations of 
q-particle groups which emerge from at least two sources. In Ref.
\cite{sark} it has further been shown that in the case of 
$S$ identical sources, each having the same charged multiplicity $n$, 
one has for $q=2$ that
\begin{equation}
D^{(k)}_{q=2}\ =\  \frac{ {n \choose 2}\,S}
                {{n \choose 2}\, S+n^2\,{S \choose2}
              }\ \
{\stackrel{n\gg 1}{\longrightarrow}}\ \ \frac{1}{S} ~.
\label{ds}
\end{equation}
This expression for $q=2$ can be generalised for $q \geq 1$  
to yield
$D_q\  \ {\stackrel{n\gg q}{\longrightarrow}}\ \
1/S\,^{q-1}\:.$
For the present work it is important to realise that 
even in the case where
the number of sources remains the same, but the condition of 
equal multiplicity is lifted, the dilution factor $D_q$
may change when the over-all multiplicity and/or its 
division between the sources is modified.\\   

In the case of two particle correlations, which obviously are genuine, 
one is able to relate \cite{cramer} the normalised cumulant $K_2$ and
the BEC correlation function $C(p_1,p_2)$, namely:
\begin{equation}
C(p_1,p_2) ~- ~1 ~= ~K_2(p_1,p_2)\ .
\end{equation}
A comparison between this expression and Eq.~(\ref{bec}) 
affords the possibility to utilise the dependence of the averaged
normalised cumulant on the hadron sources and their charged
multiplicity configuration 
to the BEC correlation 
term $\lambda e^{-r^2Q^2}$
average over the $Q$ variable. 
Specifically, let us consider the ratio of the normalised average
cumulants $K^{av(m)}_2/K^{av(k)}_2$, where the superscripts $m$ and $k$
denote two different hadron sources configurations. 
Thus one has  
$$
K^{av(m)}_2/K^{av(k)}_2  = \langle C^{(m)}(Q)-1\rangle
/\langle C^{(k)}(Q)-1\rangle =  
\int_{0}^{\infty} \lambda_m e^{-r^2_mQ^2}dQ \lll 
\int_0^{\infty}\lambda_k e^{-r^2_kQ^2}dQ =\frac{\lambda_m
r_k}{\lambda_k r_m} ~.
$$
From Eq.~(\ref{definition}) follows that $K^{av(m)}_2/K^{av(k)}_2=
D_2^{(m)}/D_2^{(k)}$ so that  
one finally obtains a simple relation between the hadron emitter radii and the
dilution factors, namely
\begin{equation}
r_m ~= ~\frac{\lambda_m}{\lambda_k}~\frac{D_2^{(k)}}{D_2^{(m)}}~r_k ~.
\label{main1}
\end{equation}

If we set the final state $k$ to be that emerging from a single
source and assign to it the value $k=1$,  
then $D_2^{(1)}$ serves as a yardstick against which dilution factors
of other final states are measured. Thus for identical sources having
the same $\lambda$ and average charged multiplicity,
Eq.~(\ref{main1}) reduces with the help of relation (\ref{ds}) to
$r_m~=~S r_1$ where $r_1$ is the emitter radius of a single hadron
source. Hence
$r$ increases linearly with $S$, the number of hadron sources. 
Although it seems a priori naturally to choose for the heavy ions reactions
the nucleon-nucleon collision as the basic single 
hadron source, the actual description of nuclei reaction 
is much more complicated
mainly due to the secondary interactions within the
nucleus. 

\section{$r_{\pi^{\pm}\pi^{\pm}}$ versus $\sqrt{s}$ in $e^+e^-$
annihilation}
\begin{figure}[ht]
\centering{\psfig{file=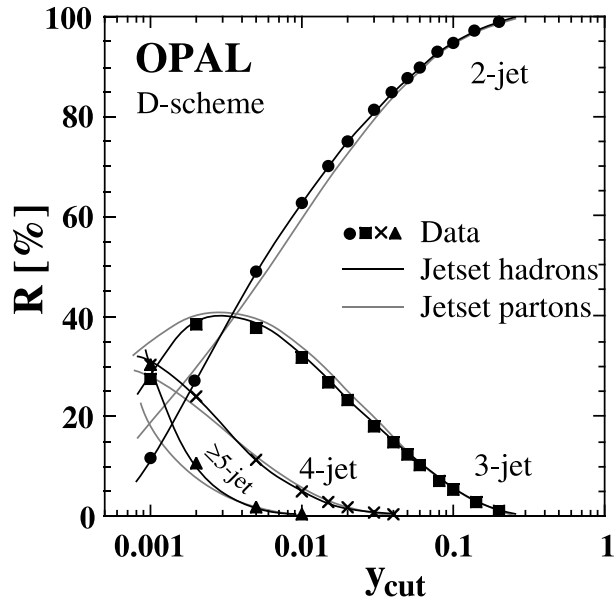,height=9cm, 
bbllx=105pt,bblly=425pt,bburx=498pt,bbury=811pt,clip=}}
\caption{\small The relative measured rate $R$ of the produced 
hadron jets in the $Z^0$ decay,
taken from reference \cite{threejets}, as a function of the jet
resolution parameter $y_{cut}$ using the Durham algorithm.
The data which is corrected for all the OPAL
detector deficiencies, but not for hadronisation effects, is compared with
two Monte Carlo models. 
}
\label{jets} 
\end{figure}

In the following we propose to study
the dependence of $r_{\pi^{\pm}\pi^{\pm}}$ on 
$\sqrt{s}$, as measured in the $e^+e^- \to$ hadrons annihilation, 
in terms of the hadron
sources given by the quark and gluon jets. Experimentally, for the grouping
of the outgoing hadrons into jets there exist several methods which differ
somewhat in their results. Among the leading ones is the known
Durham algorithm \cite{durham}, which has been applied to the
hadronic $Z^0$ decay.  The outcome of such a hadron jet analysis, as
obtained by the OPAL collaboration \cite{threejets}, 
is shown in Fig.~\ref{jets} 
as a function of the jet resolution parameter $y_{cut}$.    
As can be seen, already for values of $y_{cut} \ge 0.01$, 
the data is predominantly limited by three hadron jets.  
In the following we will assume that at $\sqrt{s} \leq  91$ GeV 
the $e^+e^- \to$ hadrons annihilation is governed by
not more than the three-jet configuration.\\ 

In the first next-to-leading process $e^+e^- \to q \bar q g$, the
differential cross section is given by (see e.g. Ref. \cite{madjid}): 
\begin{equation}
\frac{1}{\sigma}\frac{d^2\sigma}{dx_1dx_2}\ =\ 
\frac{2\alpha_s}{3\pi}\frac{x_1^2+x_2^2}{(1-x_1)(1-x_2)}\ =\
\frac{2\alpha_s}{3\pi} ~W_{qqg}(x_1,x_2) ~.
\label{xsec}
\end{equation}
Here $x_1 ~= ~2E_q/\sqrt{s}$ and $x_2 ~= ~2E_{\bar q}/\sqrt{s}$.
where $E_q$ and $E_{\bar q}$ are respectively the energies of the
outgoing quark
and anti-quark jets. 
At this point it is important to realise
that the term $W_{qqg}(x_1,x_2)$
is equivalent to the probability of finding a gluon with the energy
$E_g=\sqrt{s}-E_q-E_{\bar q}$ which here is utilised for the 
calculation of the dilution factors. 
The integration region associated to this differential cross section is defined
by $0 \leq x_1,x_2 \leq 1$ and $x_1+x_2 \ge 1$.
Further to note is that the 
expression given by Eq.~(\ref{xsec}) diverges when either the gluon is
collinear with one of the outgoing quarks or the gluon momentum
approaches zero. Some methods to handle in the integration 
these singularities are discussed e.g.
in references \cite{madjid,khoze}. Here we deal with this singularity by
introducing to the $x_1$ and $x_2$ variables an upper limit
of $x_{max}=0.99$.     
To estimate the sensitivity of our calculations to this upper
limit cutoff choice we also represent our analysis results using 
$x_{max}=0.95$.\\

For the average charged multiplicity of the quark and anti-quark jets, 
in the process $e^+e^- \to q \bar q g$,
it was found that its parametrisation is best given in terms
of an appropriate scale defined as
\cite{qaleph} 
\begin{equation}
Q_{qg}\ =\ E_q \sin\left (\frac{\theta_{qg}}{2}\right )\ \ \ {\rm{and}}\ \
\ Q_{\bar qg}\ =\ E_{\bar q} \sin\left (\frac{\theta_{\bar
qg}}{2}\right )\ , 
\end{equation}
where $E_q$ ($E_{\bar q}$) is the energy of the quark (anti-quark) jet and
$\theta_{qg}$ ($\theta_{\bar qg}$) is the opening angle
between the quark (anti-quark) jet and the gluon jet. The effective
scale for the gluon is then 
\begin{equation}
Q_g\ =\ \sqrt{Q_{qg}Q_{\bar qg}}\ .
\end{equation} 

For the dependence of the multiplicity on the scale $Q_{jet}$ we use
the parametrisation
given in Refs. \cite{madjid,qaleph}, namely:
\begin{equation}
\langle N_q(Q_{jet})\rangle ~= ~a_0 + a_1ln Q_{jet} + a_2 (ln Q_{jet})^2 
\label{nq}
\end{equation}
and
\begin{equation}
\langle N_g(Q_{jet})\rangle ~= ~R_0 + R_1\langle N_q(Q_{jet})\rangle ~,
\label{ng}
\end{equation}
where\ \
$a_0=2.74\pm 0.07,\ a_1=1.71\pm 0.05,\ a_2=-0.05\pm 0.07,\ R_0=-7.27\pm
0.52$ and $R_1=2.27\pm 0.07.$
With the help of these two parametrisations one can calculate the
2-pion
dilution factor $D_{qqg}(x_1,x_2)$, as defined in Eq.~(\ref{dilution}), 
for every set of
$x_1$, $x_2$ and $x_g=2-x_1-x_2$ values assuming for simplicity that
each jet charged multiplicity is equally divided into positive
and negative pions. Thus the average 2-pion dilution factor 
$\langle D_{qqg} \rangle$, at a given $e^+e^-$ centre-of-mass energy, 
is equal to
\begin{equation}
\langle D_{qqg} \rangle = \int_0^{x_{max}}dx_2\int_{1-x_2}^{x_{max}}D_{qqg}(x_1,x_2)
W_{qqg}(x_1,x_2)dx_1\lll\int_0^{x_{max}}dx_2
\int_{1-x_2}^{x_{max}}
W_{qqg}(x_1,x_2)dx_1 ~.
\label{average}
\end{equation}
The average dilution factors $\langle D_{qqg} \rangle$ were thus calculated
at several $\sqrt{s}$ values in the range of 10 to 91 GeV using the two 
$x_{max}$ values of 0.99 and 0.95.   
Our results for the expected
$r_{\pi^{\pm}\pi^{\pm}}$ dependence on 
energy, using Eq.~(\ref{main1}) and assuming 
a constant $\lambda$ value, are represented in Fig.~\ref{compilation}
by a band chosen to be 
normalised to the $r$ value measured by ALEPH 
\cite{alephr} at 91 GeV so as to cover also $r$ measurements at lower 
energies.
The lower and upper lines of the band, which   
correspond respectively  
to the $x_{max}$ values of 0.99 and 0.95, are seen to be very
similar in their shape. 
As can be seen, the expected dependence of $r$ on 
the energy is weak and consistent 
with the experimental observations in the energy range from
10 GeV up to the $r$ values measured at the $Z^0$ mass energy 
by the ALEPH and DELPHI 
collaborations.  
At the same time, in the frame work of the hadron source 
approach, no relation can be achieved between the reported
$r$ values of 
L3 and OPAL at $\sqrt{s}=91$ GeV  
and those
measured at lower energies.
Finally important to note is that whereas 
in the description of the jet multiplicity
via Eqs. (\ref{nq}) and (\ref{ng}) the use of $Q_{jet}$ scale is mandatory,
\cite{qaleph},
the average calculated dilution factor  
$\langle D_{qqg} \rangle$ differs, in the 10 to 91 GeV energy range,
only by about $2\%$ when the $Q_{jet}$ scale
is replaced by the jet energy $E_{jet}$. 

\subsection{The $r_{\pi^{\pm}\pi^{\pm}}$ value of the 
$Z^0$ decay into two and three jets}  
The dependence of the emitter size on the number of hadron jets in the
$Z^0$ decays has been studied
by the OPAL collaboration \cite{opalmul} which found that the $r$ value
of the 3-jet events has a value which is higher by
about $10\%$ than that obtained for the 2-jet events. From the same
study OPAL also reported that $r$ increases with the
observed charged multiplicity while the $\lambda$ parameter decreases.
\begin{figure}[ht]
\centering{\psfig{file=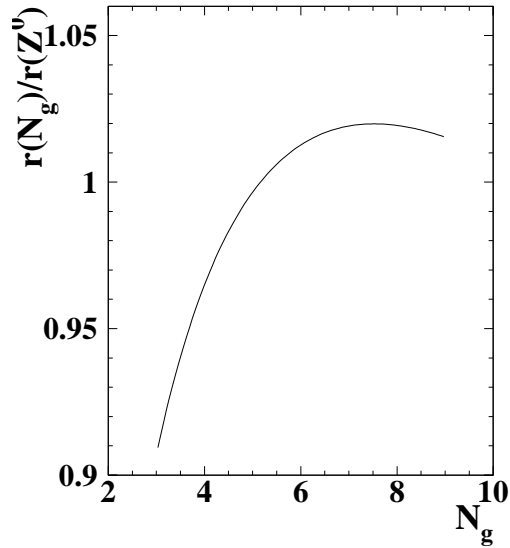,height=8cm}}
\caption{\small The expected ratio $r({\small N_g})/r({\small Z^0})$ 
of the hadronic
$Z^0$ decay as a function of $N_g$,
the gluon jet charged multiplicity, 
as calculated from the hadron source approach.}
\label{dgluon}
\end{figure}
As mentioned
before, the experimental separation between 2-jet and 3-jet events
depends on the method chosen and its parameters values
and therefore its reproduction with Eq.~(\ref{xsec}) is inaccessible.
On the other hand, 
transition from   
a 2-jet to a 3-jet $Z^0$ decay configuration
can be characterised, in the framework of the 
hadron source model, by an 
increase of $N_g$,
the number of charged
hadrons associated with the gluon jet.  
The expected ratio of 
$r({\small N_g})/r({\small Z^0})$,  
as obtained from the model using $x_{max}=0.99$, 
is shown in Fig.~\ref{dgluon} as a function of $N_g$, where 
$r({\small Z^0})$ is the
dimension value obtained from the BEC analysis of the total $Z^0$ 
final state hadrons.
As seen, this ratio increases by about $10\%$ from $N_g$ of about three  
to values above seven where the ratio settles at a level 
of about $\sim$1.02. Similar behaviour of 
$r({\small N_g})/r({\small Z^0})$ 
is also seen when the value $x_{max}=0.95$ is used in the calculations.
Since $r(Z^0)$ 
lies between the $r$ values of the 2 and 3-jet events, the
model expectation is consistent with the experimental findings.\\
 
For the comparison of the OPAL results for the $r$
dependence on multiplicity at the $Z^0$ decay  \cite{opalmul},  
with the hadronic jet
source approach one requires above all a parametrisation of the hadron
multiplicity spectrum as a function of $Q_{jet}$ 
in addition to its average
value $\langle N(Q_{jet})\rangle$. 
For the gluon jet this requirement can be achieved by using 
its hadron multiplicity distributions given in \cite{gluonjet}
for several $Q_{jet}$ values. 
As for  the quark jet,
one can obtain the needed parametrisation by using the known
average multiplicity versus $Q_{jet}$ and by assuming that
the shape of the  
multiplicity distribution is independent of the jet energy.
With these parametrisations we find that for a constant $\lambda$ 
the dimension $r$ 
indeed increases with the true 
multiplicity. 
Nevertheless, it is not surprising that we are unable to reproduce
the experimental findings in all their details and that for the 
following reasons.  
Firstly, the measured multiplicity given in
\cite{opalmul} is the observed one, which depends on the
experimental setup and analysis procedures, 
whereas the hadron jets parametrisations refer to the true
multiplicity. Secondly, the measured chaoticity $\lambda$ is found 
to decrease rather steeply with the multiplicity 
which may indicate   
an increase in the experimental data contamination as 
the number of outgoing particles increases.
Thirdly, the present study has limited itself to the case
where the decay configurations of more than three hadron jets can be
neglected. While this assumption is supported by the OPAL analysis
\cite{threejets} even at $\sqrt{s}=91$ GeV, as long as the 
BEC analysis embraces the whole final state hadrons,  
it may not be anymore valid when only the higher multiplicity part
of the hadronic $Z^0$ decay events is selected for the correlation study. 

\subsection{Quarkonium decay into hadrons}
Quarkonia, like the 
$J/\psi(3.1)$ and the $\Upsilon(9.46)$, decay into hadrons
predominantly via the three gluon ($3g$) mode, also referred
to as the direct decay mode, which leads to three hadron jets.
In addition, they also decay via the one photon, 
or vacuum polarisation, mode
which is identical in its hadron jet configuration to that of the 
$e^+e^- \to \gamma/ Z^{\star} \to q \bar q g$ background underneath.  
The $J/\psi(3.1)$ mass is too low to be analysed here in terms
of its hadron sources, mainly due to the uncertainty in the gluon and
quark jet parametrisations below $\sim 1-2$ GeV. As for the 
$\Upsilon(9.46)$, its mass may be considered just heavy enough to  
render, according to Eq.~(\ref{main1}), 
a reliable estimate of
the ratio 
\begin{equation}
\frac{r_{3g}}{r_{qqg}}\ =\  
\frac{\langle D_{qqg}\rangle}{\langle D_{3g}\rangle}~ 
\frac{\lambda_{3g}}{ 
\lambda_{qqg}}\ .
\label{ratior}
\end{equation}
To calculate the dilution factor of the three gluon-jet configuration
we consider the differential $\Upsilon(9.46)$ direct decay rate given by
\cite{koller}
\begin{equation}
\frac{1}{\Gamma_{3g}}\frac{d\Gamma}{dx_1dx_2}\ =\ 
\frac{6}{\pi^2-9}~W(x_1,x_2,x_3)~,
\end{equation}
where $x_j=2E_{g_j}/m_{\Upsilon}$. Here $W(x_1,x_2,x_3)$
is given by \cite{koller}
\begin{equation}
W(x_1,x_2,x_3)\ =\ \frac{1}{x_1^2 x_2^2 x_3^2}
~[x_1^2(1-x_1)^2+x_2^2(1-x_2)^2+x_3^2(1-x_3)^2]
\end{equation}
with the condition that $x_1+x_2+x_3=2$. 
Whereas for the calculation of the average dilution factor 
$\langle D_{qqg}\rangle$ one has a well defined
$Q_{jet}$ scale for the quarks and gluon jets 
as formulated in Eqs. (\ref{nq}) and (\ref{ng}), not so for the 
three gluon configuration. However, as mentioned before, the average
dilution factors do change only by $\sim~2\%$ when the $Q_{jet}$ scale is 
replaced by $E_{jet}$. This being the case we proceeded to calculated
$r_{3g}/r_{qqg}$ using the jet energy $E_{jet}$ rather than the
$Q_{jet}$ scale while incorporating a 2$\%$ uncertainty into our final result.
The evaluation of 
the average dilution factor $\langle D_{3g} \rangle$ is then carried
out via Eq.~(\ref{average}) where  
$D_{qqg}(x_1,x_2)$ and $W_{qqg}(x_1,x_2)$
are replaced respectively by $D_{3g}(x_1,x_2)$
and $W_{3g}(x_1,x_2)$. From these calculations and those
carried out to obtain $\langle D_{qqg} \rangle$ at 
$\sqrt{s}=9.46$ GeV, we find that
the ratio $\langle D_{qqg}\rangle /\langle D_{3g}\rangle$ 
is lying in the range
of 1.06 to 1.11 corresponding to the integration
cutoff values $x_{max}=$ 0.99 and 0.95.\\

An early BEC analysis of the CLEO collaboration \cite{cleo} yielded 
the values of  
$r(\Upsilon \to 3g)=0.99\pm 0.14$ fm and
$r(\Upsilon \to qqg)=0.86\pm 0.15$ fm so that their ratio
is equal to $1.15\pm 0.26$.
A more recent BEC measurements of 
$e^+e^- \to$ hadrons at the $\Upsilon$ mass energy,
described in Ref. \cite{blinov}, quotes 
the values $r(\Upsilon \to 3g)=0.73\pm0.10\pm0.04$ fm 
and $r(\Upsilon \to qqg)=0.83\pm 0.22\pm0.05$ fm, from which follows 
that $r(\Upsilon \to 3g)/r(\Upsilon \to qqg)=0.88\pm0.28$. 
To compare these results with the expectation of 
the hadron source approach one needs to know the true 
$\lambda_{3g}$ and $\lambda_{qqg}$ values. 
However as mentioned before, these are difficult
to estimate due to the fact that
they are sensitive to the purity level of the experimental data sample.
Inasmuch that 
$\lambda_{3g} \approx \lambda_{qqg}$, an assumption supported by the
analysis results reported in
\cite{blinov}, the two independent measurements of 
$r_{3g}/r_{qqg}$ are consistent
within one standard deviation with the expectation of the hadron
source approach.     

\section{Summary}
The dependence of the Bose-Einstein correlation of two identical
bosons, produced in
$e^+e^-$ annihilations, on the configuration of the hadron sources 
has been taken over from the known  
cumulants dependence on these sources. Assuming that
correlation exists only for hadrons
emerging from the same source, leads to a BEC dilution
whenever the hadron-pair emerges from two different sources.   
A relation between the amount of dilution and the BEC extracted 
dimension $r$ is formulated. Specifically, we identify in the
$e^+e^-$ annihilation the outgoing particle jets as the hadron sources 
which up to centre-of-mass energy of 
$\sqrt{s}=91$ GeV are essentially
limited to three.\\ 

The comparison of the hadron source approach to the experimental
findings is somewhat hindered by
the need to know the genuine values of the chaoticity 
$\lambda$ parameters, and in some of the cases also
the distribution of the true multiplicity, both as a function of the jet
energy. In addition, 
the extension of the hadron source approach to $e^+e^-$ annihilations at
energies above 91 GeV seems to
require the inclusion of configurations higher than three hadron jet sources
which are not dealt with here.\\

Insofar that $\lambda$ remains the same over the whole energy range
from 10 to 91 GeV,    
the hadron source approach expects the BEC extracted
$r$ value to be only slightly dependent on energy
in accordance with the experimental observations. 
As for the hadronic $Z^0$ decay, the seen rise of $r$ by some $10\%$ 
when shifting from 
the two
to the three hadron-jet configuration, 
is also expected from
the hadron source approach. Finally the model is consistent
within errors with the
experimental findings for the ratio between the $r$ values obtained for  
the three gluons and to the one photon hadronic decay 
modes of the $\Upsilon(9.46)$.

\vspace{4mm}

\noindent
{\bf Acknowledgements}\\
I would like to thank Y. Achiman, 
G. Bella, E. Reinherz-Aronis and E.K.G. Sarkisyan for many
helpful discussions.
\vspace{3mm}\\
 
{\small
\linespread{1.0}

}
\end{document}